*sensors*

MDPI

Article# A Telerehabilitation System for the Selection, Evaluation and Remote Management of Therapies

**David Anton** [1,*], **Idoia Berges** [2], **Jesús Bermúdez** [2], **Alfredo Goñi** [2] **and Arantza Illarramendi** [2]

1. Department of Electrical Engineering & Computer Sciences, University of California, Berkeley, CA 94720, USA
2. Department of Languages and Information Systems, University of the Basque Country UPV/EHU, 20018 Donostia-San Sebastián, Spain; idoia.berges@ehu.eus (I.B.); jesus.bermudez@ehu.eus (J.B.); alfredo@ehu.eus (A.G.); a.illarramendi@ehu.eus (A.I.)
* Correspondence: davidantonsaez@eecs.berkeley.edu; Tel.: +34-943-015-109Received: 16 March 2018; Accepted: 4 May 2018; Published: 8 May 2018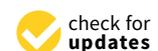

**Abstract:** Telerehabilitation systems that support physical therapy sessions anywhere can help save healthcare costs while also improving the quality of life of the users that need rehabilitation. The main contribution of this paper is to present, as a whole, all the features supported by the innovative Kinect-based Telerehabilitation System (KiReS). In addition to the functionalities provided by current systems, it handles two new ones that could be incorporated into them, in order to give a step forward towards a new generation of telerehabilitation systems. The knowledge extraction functionality handles knowledge about the physical therapy record of patients and treatment protocols described in an ontology, named TRHONT, to select the adequate exercises for the rehabilitation of patients. The teleimmersion functionality provides a convenient, effective and user-friendly experience when performing the telerehabilitation, through a two-way real-time multimedia communication. The ontology contains about 2300 classes and 100 properties, and the system allows a reliable transmission of Kinect video depth, audio and skeleton data, being able to adapt to various network conditions. Moreover, the system has been tested with patients who suffered from shoulder disorders or total hip replacement.

**Keywords:** telerehabilitation; virtual therapy; Kinect; eHealth; telemedicine## 1. Introduction

Traditional rehabilitation takes place in rehabilitation centers or hospitals, which requires that patients travel to their appointments. This travel is often associated with both time and financial costs [1]. An alternative rehabilitation method involves using telerehabilitation technologies, which allow rehabilitation services to be delivered directly to patients' homes [2]. Telerehabilitation systems have the potential of providing anywhere and anytime physiotherapy support for different groups of persons such as the elderly, disabled and sick, facilitating their contact with caregivers and improving their quality of life. Several studies indicate the therapeutic usefulness of telerehabilitation systems [3,4]; and tests based on virtual interaction have shown that these can be as effective as traditional treatments [5,6]. In addition, as the abandonment of classical rehabilitation sessions because of boredom or disinterest is relatively frequent, the motivating character of telerehabilitation systems is an important factor to consider. In this sense, several studies have found that Virtual Reality (VR) game-based telerehabilitation is perceived as enjoyable and engaging and that it can increase the intensity of rehabilitation and the patient's enjoyment [7–9]. Another advantage of telerehabilitation programs is the easy access of the healthcare professionals to the data collected from patients via the Internet and mobile devices [10–12]. Data collected via sensors during telerehabilitation sessions can

*Sensors* **2018**, *18*, 1459; doi:10.3390/s18051459   www.mdpi.com/journal/sensors

be further processed to provide more effective health interventions [13,14]. Finally, telerehabilitation is significantly more time-efficient for both physiotherapists and patients, even when travel time to regular therapies is excluded [15].

A basic telerehabilitation system has at least one camera that allows a physiotherapist to see the patient and monitor the therapy directly (videoconferencing). More complex systems include sensors that can record the movements of the patient. Existing telerehabilitation systems are oriented toward the treatment of many pathologies. In a first approximation, they can be classified into three main groups. The first group comprises those systems that propose that users wear devices, such as [16–19]. Thus, the system for task-oriented games presented in [16] evaluates whether people with cognitive impairment can reach some predefined locations. The system presented in [17] uses smartphones' built-in inertial sensors to monitor exercise execution and to provide acoustic feedback on exercise performance and execution errors. The systems presented in [18,19] make use of sensorized garments and sensors, respectively, that are worn by patients for evaluating a series of exercises related to the upper limbs.

The second group includes those systems that advocate that users do not wear devices, but use low-cost non-intrusive tracking devices such as the Nintendo Wii Remote, Leap Motion or Kinect. The system based on the Nintendo Wii Remote presented in [20] uses an accelerometer to record in 3D and focuses on rehabilitation exercises of the upper limbs. The system presented in [21] uses a webcam and adaptive gaming for tracking finger and hand movement. Trackers are attached to some objects, and a webcam captures the patient's hand to generate some metrics that provide information about the quality, efficiency and skill of the patient. The system presented in [22] uses a Leap Motion device to conduct a video game-based therapy that evaluates the hand's ability and grasp force. Furthermore, Kinect has become one of the most widely-used tracking devices in telerehabilitation [23–36]. The device offers visual tracking without markers, which allows users to control and interact with applications. The data provided can be used to analyze movements, gestures and body postures and can assist in obtaining scores for medical analysis [23]. Among the Kinect-based systems that consider different pathologies, we can mention the following ones: Kinerehab [24], an occupational therapy system where patients can perform three different exercises: lift arms front, lift arms sides and lift arms up; a game aimed at training a dynamic postural control system [25] for people with Parkinson's disease; a 21-game prototype system [26] that evaluates upper body exercises for individuals with spinal cord injury; an upper limb rehabilitation system [27] for stroke survivors, and a similar system, but for people with cerebral palsy [28]; a full body gait analysis system [29]; and finally [30], where fine motor movements are evaluated (like hand and wrist movement) in patients with traumatic brain injury. In addition to the previous works, we would like to mention the system presented in [31], which explores the combined use of Kinect and inertial sensors in order to provide robust hand position tracking; and the Tele-MFAsT telerehabilitation system [32], which has been designed for remote motion and function assessment that facilitates streaming and visualization of data (video, depth, audio and skeleton data) from remotely-connected Microsoft Kinect devices. There are also some commercial solutions that make use of Kinect, such as [33–36], which allow physiotherapists to customize and monitor the therapy sessions of the patients and analyze their evolution. Nevertheless, they do not support both of the two innovative functionalities supported by the Kinect-based Telerehabilitation System (KiReS)–automatic recommendation of therapies and two-way teleimmersion—which we have designed, implemented and the interest in which has been tested.

Finally, in the third group, proposals that belong to the telerehabilitation robotics area can be found. According to [37], those systems are considered as cost-effective alternatives compared to clinic-based therapy, but still, some barriers need to be addressed in order to get a more widespread acceptance. In any case, clinical guidelines recommend these kinds of systems for the recovery of the lost functions in some pathologies such as acute/subacute or chronic strokes. For instance, we can mention the following: MOTORE++ [38], a rehabilitation robot that restores upper limb functionality by using a rolling device; HOMEREHAB [39], which tries to help people who suffer from hemiparesis regain movement of their weak arms and legs by making use of a floor-grounded haptic interface; a WAM robot [40], which is used for upper extremity rehabilitation of patients whose legs can move and which follows the patient trunk movement tracked by Kinect in real time; and those in [41,42], which use haptic sticks in a virtual environment with rehabilitation games for upper limb movement therapy and assessment.

The development of a telerehabilitation system requires interdisciplinary collaboration in order to achieve a good result. Thus, in addition to software and computer engineers for designing, modeling and implementing the system, the presence of experts in the field of rehabilitation, such as doctors and physiotherapists, is required. The role of the end users as active participants must also be taken into account during the whole process of the design, testing and deployment of new technologies. In this sense, we believe that for a telerehabilitation system to be considered as useful, it must be able to help physiotherapists in performing the following tasks: (1) selecting appropriated therapies for patients; (2) evaluating the therapies performed by the patients; and (3) managing those therapies in a remote way. The telerehabilitation system must also empower the patients in following their therapies by motivating them, so that they do not abandon them, as well as by providing the patients with feedback that allows an autonomous evaluation without the direct intervention of the physiotherapist during rehabilitation sessions.

The main emphasis of the majority of works previously mentioned is the evaluation of the therapies' tasks, and they only consider superficially other tasks such as the selection of adequate therapies. For this reason, we have built KiReS (Kinect TeleRehabilitation System), an innovative system that provides a solution for different tasks. KiReS can be a very useful system for both physiotherapists and patients. In the first case, KiReS can accompany them through the entire process of managing the rehabilitation of the patients (designing and selecting therapies, following patients' progress, communicating with patients in virtual environments). In the second case, KiReS can accompany patients in their rehabilitation process by motivating and informing them about their progress. In order to build our system, we have collaborated with physiotherapists to add the adequate expert knowledge to the system and with patients to validate their interest.

The rest of the paper is organized as follows. In Section 2, the main features of KiReS are shown: the workflow of activities, the modules of its architecture that support those activities and the specific methods implemented in those modules. In Section 3, the main results obtained from two real trials are described. Finally, in Sections 4 and 5, some discussion and conclusions are presented, respectively.

## 2. Materials and Methods

In this section, first, we show the main activities associated with the use of KiReS through a workflow represented using a UMLdiagram and, then, the architecture of the system that supports those activities. Next, the main methods applied in KiReS for selecting, evaluating and remotely managing therapies are explained briefly.

### 2.1. KiReS Workflow

The use of KiReS involves performing the activities shown in the UML activity diagram of Figure 1, which are executed by two human actors (physiotherapists and the patients) and by the data analyzer, which is a system actor that performs the data analytics processes.

With respect to the therapy selection, physiotherapists have to assign rehabilitation exercises to the patients (Assign Exercises). In order to do this, the exercises must have been previously created (Create New Exercises), along with adequate tests to obtain feedback from the patients (Create Tests). With respect to therapy evaluation, patients have to perform the assigned exercises, which are automatically monitored by KiReS (Perform Monitoring Exercises), and answer to the evaluation tests (Answer Rehabilitation Tests). After that, the data analyzer performs the data analysis with all the data generated during the patients' rehabilitation sessions (Data Analytics).

The result of that data analytics is provided to the physiotherapists so that they can evaluate the evolution of the patients (Evaluate Data Analytics Results). With respect to the remote management of therapies, depending on the evolution of the patients, physiotherapists can choose to establish remote rehabilitation sessions in real time with the patients (Teleimmersion session). In addition to the activities, there are two important objects in this activity diagram: (a) the KiReSdb database, which stores the recorded rehabilitation exercises and tests, as well as the data generated by the patients while they perform the exercises and the answers that they give to the rehabilitation tests; and (b) the TRHONT ontology, which is used to help physiotherapists assign exercises and evaluate the results of the data analytics obtained from the data stored in KiReSdb.

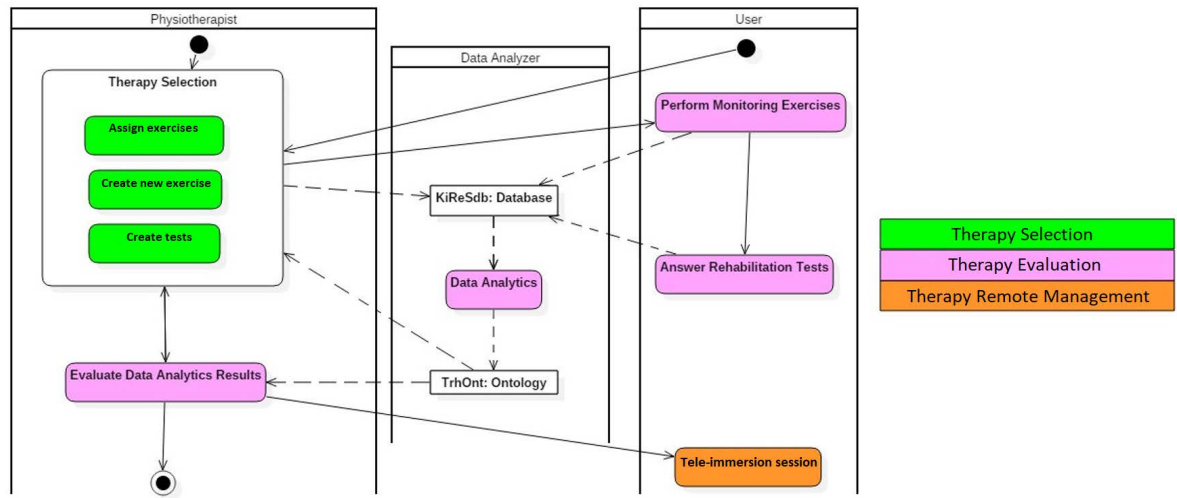

**Figure 1.** The Kinect-based Telerehabilitation System (KiReS) activity diagram.

## 2.2. KiReS Architecture

The architecture of KiReS (see Figure 2) is composed of three modules that support the main activities provided for the human actors (patients and physiotherapists) and the system actor (data analyzer): the Interface and Communications Module, the Recognition and Evaluation Module and the Knowledge Extraction Module. These modules implement the specific methods that are going to be explained in the following subsections, through which the activities defined in the KiReS workflow can be performed.

The Interface and Communications Module provides patients with friendly and helpful interfaces that include motivational features such as avatars in order to perform exercises defined in therapies or treatments. This module also provides physiotherapists with interfaces to define and create new exercises that can be included in therapies. Lastly, it also allows real-time teleimmersion sessions among patients and physiotherapists if required and is responsible for managing and transmitting all the data generated in those sessions.

The Recognition and Evaluation Module is responsible for monitoring the patients when they are performing the exercises defined in their therapies in front of the Kinect device. The module evaluates whether the patients are executing those exercises properly, by comparing them with the reference exercises recorded previously by the physiotherapists. The evaluation is made by an exercise recognition algorithm, which is applied to the data of the skeleton joints captured in real time by the Kinect. As a result, patients can receive appropriated feedback about how they are doing (through the aforementioned Interface and Communications Module).

The Knowledge Extraction Module is involved in the therapy selection process performed by physiotherapists. This module relies on the TRHONT ontology, which is used for reasoning and which is enriched with new knowledge extracted from the data generated during physiotherapy sessions (by the Recognition and Evaluation Module). Physiotherapists can also reevaluate patients and modify their therapies depending on such new generated knowledge.

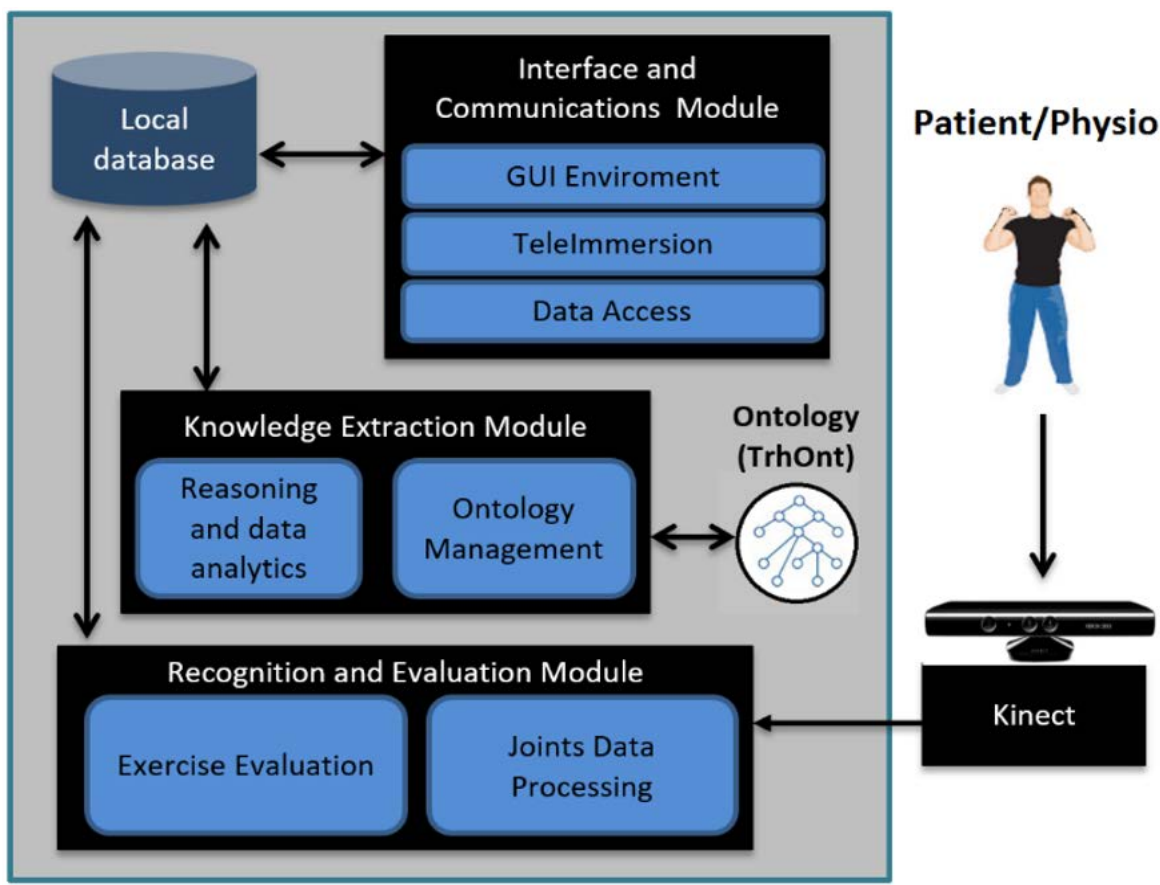

**Figure 2.** System architecture.

*2.3. Therapy Selection Methods*

In order to accomplish the activities associated with the therapy selection process, three methods are needed: Create New Exercise, Create Tests and Assign Exercises.

2.3.1. Create New Exercise

KiReS offers an interface for the physiotherapist that provides assistance for creating exercises step by step. Exercises can be created from scratch or can be reused if they are already recorded in the KiReS database. This interface is handled by the Interface and Communications Module of the system architecture (see Figure 2). In the exercise model used in KiReS, a body posture is the simplest element that composes an exercise and therefore necessary for the definition of any other structure. The physiotherapist performs the posture in front of the Kinect, and the system records it. Movements have an associated name to identify them and are defined with two postures (initial and final) and with the recording of the transition between those postures. The relevant joints that best represent the transition from the initial posture to the final posture are selected, recorded and stored. Data representation for the information concerning the name, the initial and final postures, the type of movement (e.g., flexion), the joint and the range of motion involved is added to the ontology to allow reasoning over movements.

Lastly, exercises are defined by assigning movements to them. Simple exercises can consist of just one movement, but complex exercises are a combination of basic movements that form a sequence. The exercise creation interface (see Figure 3) allows the physiotherapist to define the composition of an exercise. It shows a form to fulfill data about the exercise and two lists, one with the movements that have been assigned to that exercise and the other with the movements that are available to be added. Once stored in the system (in the database and in the ontology), the exercise will be available to be included in a therapy session.

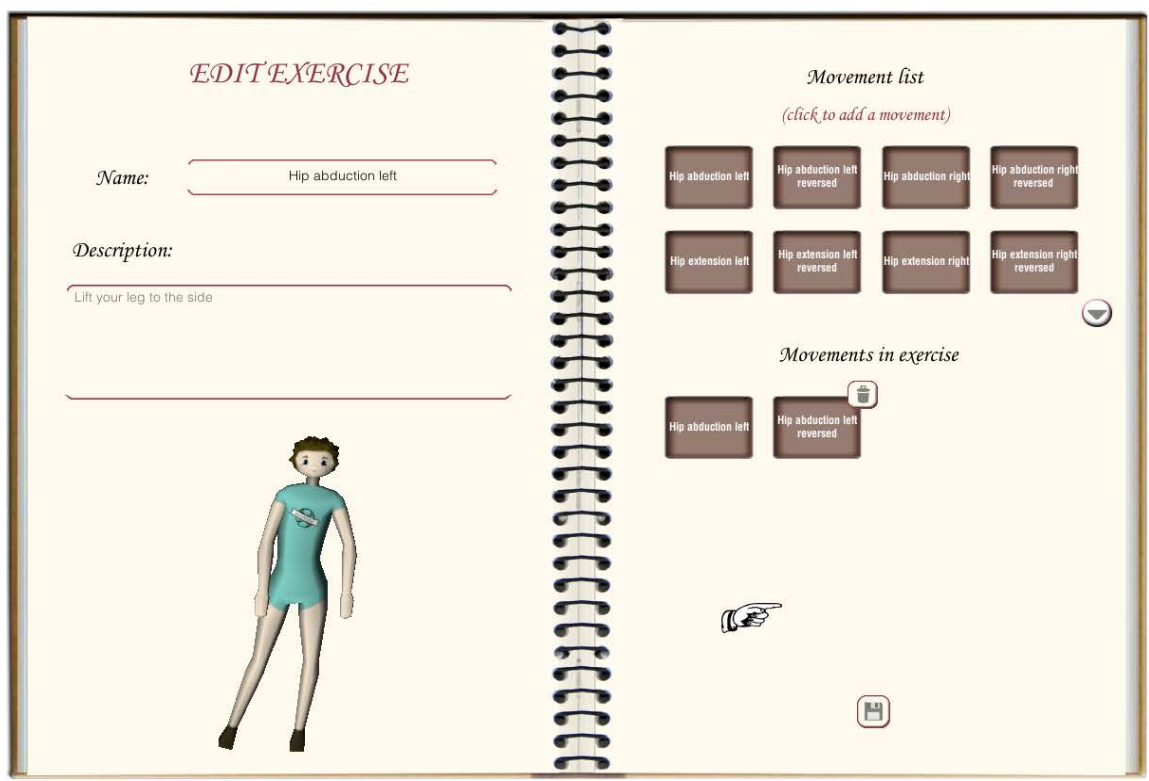

**Figure 3.** Interface for the physiotherapist.

2.3.2. Create Tests

Performance evaluation is an important factor in a therapy session. In the specialized literature, many user-oriented tests can be found. This kind of test is designed to be answered by the user after ending a therapy session. The answers of the user provide qualitative and quantitative information about their state. Answers to questions about daily life or pain suffered can provide useful feedback to the physiotherapist as a complement to the objective information that is automatically retrieved during exercise execution. Since these tests are widely used in physiotherapy sessions, we decided to incorporate in KiReS the functionality that supports them.

The Interface and Communications Module of the system architecture also provides the physiotherapists with assistance to create these tests. Our proposal includes the option of adding two types of subjective evaluation tests: auto-tests and the Visual Analogue Scale (VAS).

The auto-test interface is oriented to create, manage and evaluate auto tests. These auto tests include questions about different aspects of users' daily life, and the possible answers are valued differently depending on their severity. The tool to manage these tests lets the physiotherapist define the questions of the test and the possible answers with their scores. By default, the tests are evaluated by adding the scores of the provided answers and giving a final result, but the tool also allows the physiotherapist to define the type of function to be applied to the scores. For example, the system can count the number of answers with a certain score or give the result as a percentage depending on a fixed value. Once a test is defined, the physiotherapist can assign it to a therapy, so that the user will have to answer the test after ending a session.

Another evaluation tool used in physiotherapy that we have incorporated with KiReS is the Visual Analogue Scale (VAS). The VAS is a technique used to measure subjective phenomena like pain. It is a self-reporting device consisting of a line of a predetermined length that separates extreme boundaries of the phenomenon being measured [43]. The user sees the image A, on which they mark a point on the line between the "no pain" label and the "worst pain ever" label (see Figure 4). This datum is incorporated with the ontology and can be accessed by the physiotherapist for its analysis. As in the auto tests, the physiotherapist decides when to present the test to the user.

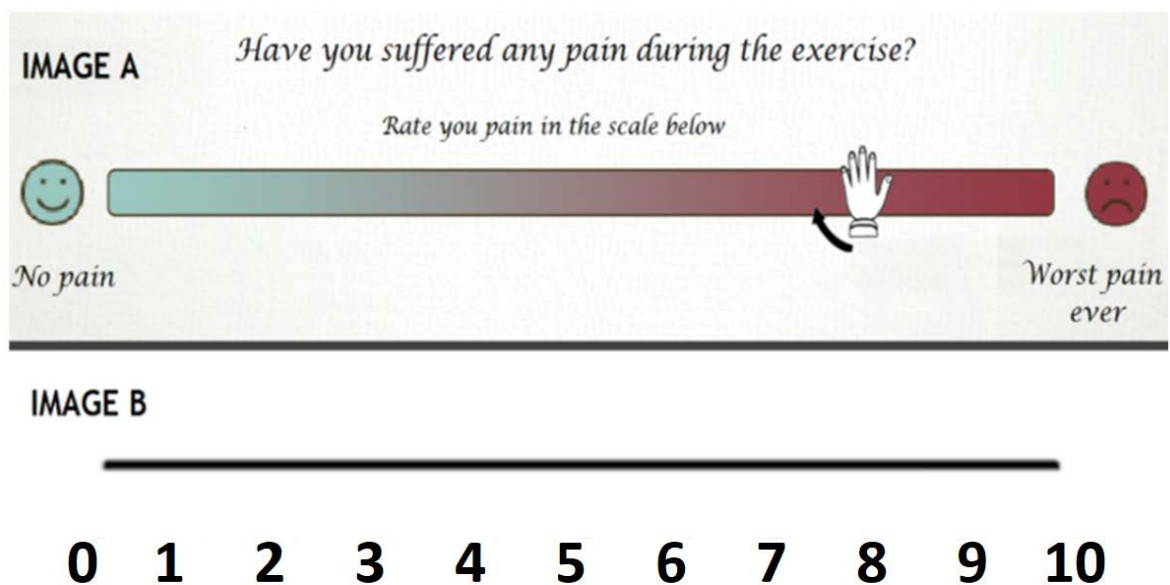

**Figure 4.** Visual Analogue Scale (VAS) example.

2.3.3. Assign Exercises

During the first visit to the physiotherapist, patients are physically evaluated and their patient record is created. Overall, the patient's record consists of information about personal and family data, symptoms, results of physical examination (joint movement), diagnoses, reported pain value and the recovering goals that are pursued, which cover several relevant aspects for the therapy. The patient's record will change over time as the patient advances in the therapy. Joint movement exploration information will be updated after each session with the data gathered from the Kinect to keep a record of the patient's evolution. Moreover, the information contained in the patient record is included in the TRHONT ontology. This ontology assists physiotherapists in recording and searching for information about the physical therapy record of a patient; identifying in which phase of a treatment protocol a patient is; and identifying which exercises are most suitable for a patient at some specific moment. Ontology reasoning plays a crucial role in these tasks. The TRHONT ontology is based on the Foundational Model of Anatomy (FMA) ontology [44] and is composed of four interrelated types of knowledge (personal data of patients, anatomy, movements and exercises and expert's domain). The components of this ontology have been designed and validated by collaborator physiotherapists to create a tool with the most relevant concepts for physiotherapy and therapy planning. The functionality to manage the TRHONT ontology is provided by the Knowledge Extraction Module of the system architecture (see Figure 2).

For example, in Figure 5, we show part of the knowledge stored for a fictional patient named John that has gone through a Total Hip Replacement (THR) surgery.

In general, rehabilitation therapies follow protocols that contain recommended exercises for a pathology classified in phases (e.g., see Table 1). Each phase contains the exercises to be performed, as well as the conditions that indicate when a patient is in that phase. These conditions are indicated, for example, in terms of the Range Of Motion (ROM) that patients achieve and the pain they report using a Visual Analogue Scale (VAS). Notice that exercises valid at any phase are appropriated also for subsequent phases: the physiotherapist can select them, for example, in order to warm the joint up.

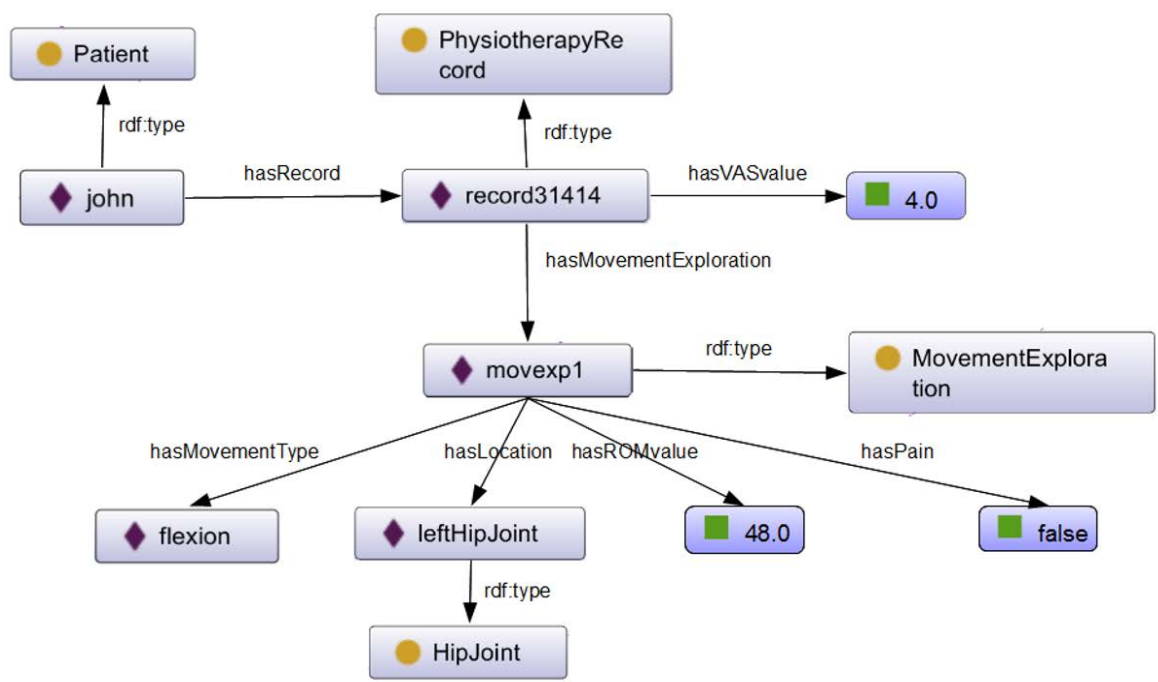

**Figure 5.** Knowledge excerpt about John contained in TRHONT.

**Table 1.** Conditions for the first two phases of a general protocol for Total Hip Replacement (THR protocol). ROM, Range Of Motion.

| Phase | Recommended Exercises | Patient Classification |
|---|---|---|
| I | Hip Flexion ROM 60<br>Hip Abduction ROM 20<br>Hip Extension ROM 20<br>Soft Squats | (ROM Flexion < 60 OR ROM Abduction < 20 OR ROM Extension < 20) AND<br>Moderate General Pain (VAS $\leq$ 6) |
| II | Hip Flexion ROM 90<br>Hip Abduction ROM 20<br>Hip Extension ROM 30<br>Soft Squats | (60 $\leq$ ROM Flexion AND 20 $\leq$ ROM Extension)<br>AND (ROM Flexion < 90 OR<br>ROM Abduction < 20 OR<br>ROM Extension < 30)<br>AND Mild General Pain (VAS $\leq$ 2) |

The TRHONT ontology contains logical axioms for the description of treatment protocols, including their phases, exercises and movements that compose the exercises. For example, in Figure 6, we present the description of the 'Hip Flexion ROM 60' exercise recommended in Phase I of the THR protocol. This refers to exercises with flexion movements with a maximum ROM of 60.

Since the THR surgery protocol specifies that hip flexion exercises with movements up to 60° are suitable for patients in Phase I of that protocol, any specific exercise included in the TRHONT ontology that complies with the definition of `HipFlexionROM60` (e.g., a hip flexion with maximum ROM of 40) will be automatically classified as an exercise for that phase (`ExerPhase1THR`) and will be recommended for patients who are in Phase I of the protocol (represented by `PatientPhase1THR`).

In the case of John, a patient that had THR surgery, a reasoning process with the TRHONT ontology over the information known about him will classify him in Phase I of the THR protocol (see Figure 5), because he has reported a VAS value of 4.0 and has obtained an ROM value of 48.0 in the flexion movement exploration of his left hip joint. Therefore, KiReS will suggest as recommended a set of exercises described in the TRHONT ontology that comply with the types of exercises of Phase I of Table 1. Moreover, the system is flexible enough to allow the physiotherapist to specify rules for certain patients that will generate different sets of recommended/contraindicated exercises (e.g., 'John can perform hip extension exercises with ROM up to 25°'). As a result, a possible therapy plan for John could be the one in Figure 7. More details about the TRHONT ontology, which contains more than 2300 classes and 100 properties, can be found in [45].

$$
\begin{aligned}
\texttt{HipFlexionROM60} &\equiv \texttt{Exercise } and \texttt{ hasMovement } only \texttt{ MovHipFlexionMax60} \\
\texttt{MovHipFlexionMax60} &\equiv \texttt{Movement } and \texttt{ hasComponent } some \texttt{ (Submovement} \\
&\quad and \texttt{ hasLocation } some \texttt{ HipJoint} \\
&\quad and \texttt{ hasMovementType } some \texttt{ Flexion} \\
&\quad and \texttt{ hasROM.value}[<= 60]) \\
\texttt{HipFlexionROM60} &\sqsubseteq \texttt{ExerPhase1THR} \\
\texttt{PatientPhase1THR} &\sqsubseteq \texttt{recommended } some \texttt{ ExerPhase1THR}
\end{aligned}
$$

**Figure 6.** Axioms for ontology classes `HipFlexionROM60` and `MovHipFlexionMax60`.

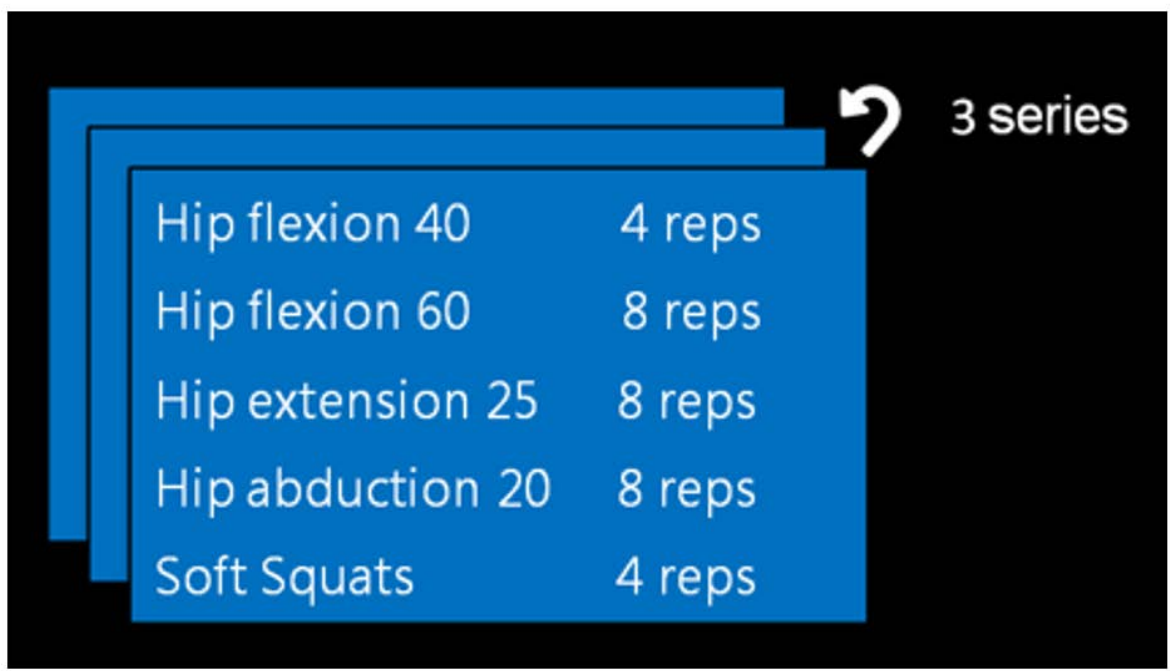

**Figure 7.** Therapy plan for John.

*2.4. Therapy Evaluation Methods*

In order to accomplish the activities associated with the therapy evaluation process, two methods are needed: Perform Monitoring Exercises and Data Analytics.

2.4.1. Perform Monitoring Exercises

The Interface and Communication Module of KiReS offers an interface (see Figure 8) that provides a game-like immersive experience that motivates and makes the therapy more enjoyable.

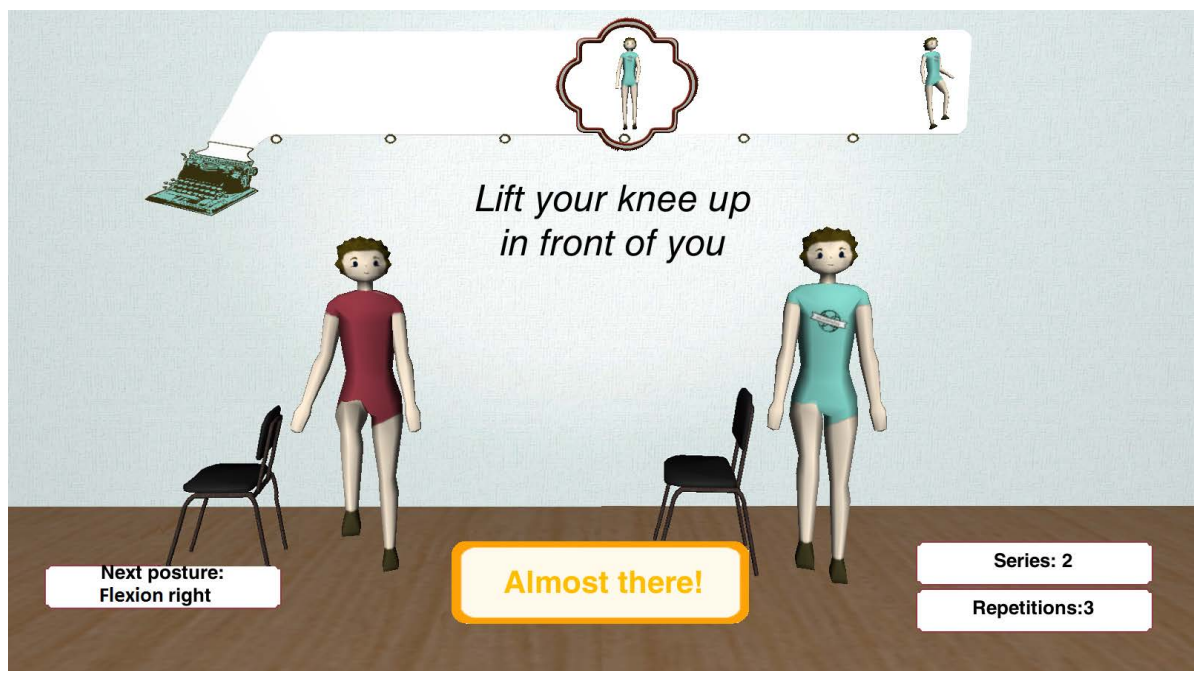

**Figure 8.** User interface.

The interface presents two 3D avatars to guide the patient. The avatar on the left provides guidance, by showing the posture the patient has to reach or the movement the patient has to do. The avatar on the right shows in real time the movement that the patient is performing. The interface also includes informative boxes at the bottom that provide information about the ongoing therapy session to the patient. The two boxes on the right show the number of series and repetitions left. When the patient has completed all the series, a session is finished. The box on the left shows the name of the next posture the patient has to reach. The box in the middle shows the 'state' of the current movement, and it is continuously updated by the exercise recognition algorithm in real time. Besides, when the patient is close to reaching a posture, the box indicates with a three-level color scale (red, yellow and green) how close they are from reaching the posture. In the upper center of the screen, there is a ribbon that shows the exercise as a list of postures that have to be reached in the current execution. This ribbon is updated as the patient completes exercises in order to show in every moment how many are left. Under this ribbon, a textual explanation of the exercise is displayed. In summary, the interface through the avatars and the boxes gives real-time information to the patient. This way, the system empowers and keeps the patients aware of their therapy.

The Recognition and Evaluation Module of the system architecture is responsible for monitoring the performance of the exercises included in the therapies assigned to the patients. It provides the implementation of the exercise recognition algorithm. An exercise is composed of a series of movements. These movements are characterized by an initial posture, a final posture and the angular trajectories of the limbs that are involved in the movement between the initial and final postures. For identifying each of the initial and final postures, a posture descriptor of 30 features (18 binary features that give information about the relative position in 3D of some joints and 12 features that represent the angles formed by the different limbs of the body) is generated. These descriptors are obtained from the skeleton structure provided by Kinect. Then, the new descriptors are classified by comparing them to previously-recorded and annotated posture descriptors. If the distance is less than a threshold value, the corresponding class is assigned. To represent the movement between both postures, the sequence of angular values of the limbs that are in a different position from the initial posture to the final posture is captured. The sequence is then compared to the previously-stored trajectory for that movement, and again, a similarity value is obtained. Finally, the algorithm analyzes the results of the performance of the exercise by taking into account the similarity values obtained in the previous step (see more details in [46]). The accuracy of the algorithm is around 94%.

2.4.2. Data Analytics

Once exercises are performed by patients, KiReS provides actionable information to physiotherapists and patients through the two submodules of the Knowledge Extraction module: the Ontology Management submodule and the Reasoning and Data Analytics submodule. Within the first submodule, the aforementioned TRHONT ontology assists physiotherapists in their daily tasks via reasoning supported by semantic technology.

Concerning data analytical functionalities, by analyzing the data retrieved from Kinect, the Reasoning and Data Analytics submodule of KiReS provides insight into a number of quantitative and qualitative measures (e.g., posture rating, exercise rating, balance, etc.) that can be useful for the physiotherapist to customize and adapt the patient's therapy and for the patients to be aware of their improvements. If the patient is not discharged, the physiotherapist will use this new information to modify the therapy.

For instance, KiReS allows the comparison of the results of one patient with the results obtained by other patients that fulfill some conditions. For example, in Figure 9, the exercise rating given by KiReS to several patients during the sessions they performed and the corresponding average rating compared to John's results (in blue) can be observed. In Figure 10, the performance, measured as the accuracy of achieving postures, can be found.

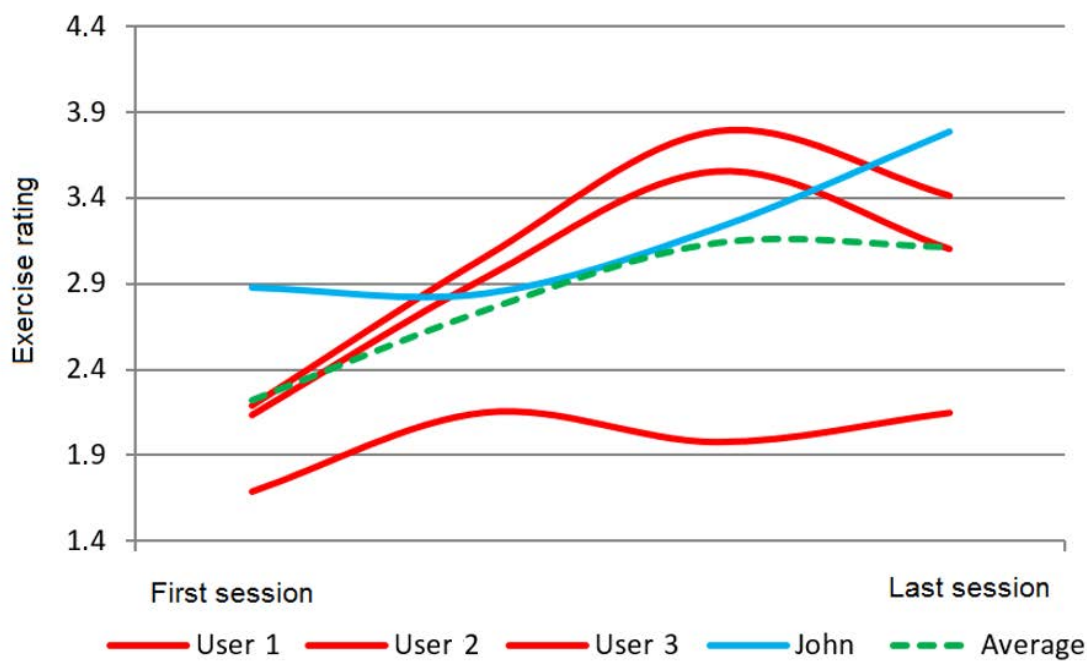

**Figure 9.** Performance over time (exercise rating).

Furthermore, other types of analysis are possible; for example, the analysis of the maximum, minimum and arc ranges the patient has achieved during shoulder exercises can be discovered by using the raw data of the body joints recorded with the Kinect on several executions of an exercise.

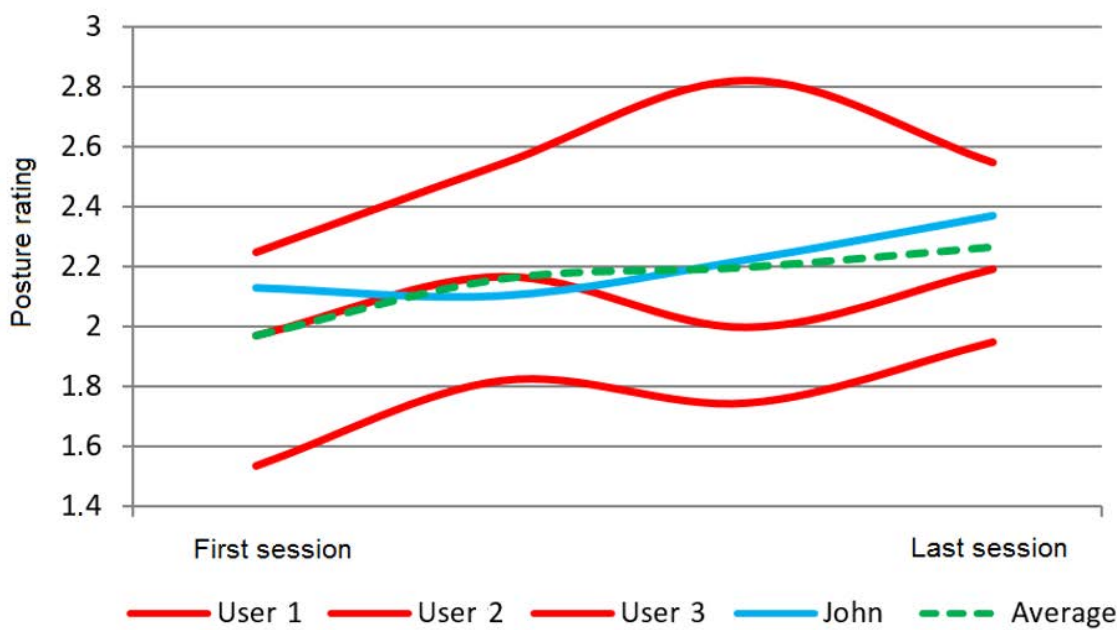

**Figure 10.** Performance over time (posture rating).

*2.5. Therapy Remote Management Method*

The only method required for the remote management of therapies is the teleimmersion session. This method, implemented in the Teleimmersion submodule of the Interfaces and Communications Module, allows a two-way real-time multimedia communication and interaction between two remote users (physiotherapist and patient) inside a virtual environment. This provides a convenient, effective and user-friendly experience when performing the telerehabilitation sessions. The submodule relies on a communication framework (called KinectRTC), based on WebRTC and the first version of Kinect.

Web Real-Time Communication (WebRTC) is an API (Application Programming Interface) that enables peer-to-peer audio, video and data sharing between peers in real time. WebRTC alleviates some of the issues in multimedia communication between various platforms and across different network configurations [47,48], and it manages congestion, data synchronization and multimedia buffering. Therefore, it has been widely adopted for video conferencing solutions and also integrated across several web browsers as a communication standard. Moreover, the implementation of secure communication protocols and platform independency makes WebRTC an ideal network framework for personal data, medical data and real-time interaction in remote locations, as all WebRTC components require mandatory encryption [49].

KinectRTC allows for real-time interaction between a physiotherapist and a patient inside a virtual environment, while also providing quantitative information on the patient's movement. It facilitates stable and secure transmission of video, audio and Kinect data (i.e., camera parameters, skeleton data and depth image) in real time between two sites. More precisely, KinectRTC relays on a P2P architecture consisting of three main modules. The Peer Connection Management Module is in charge of managing the connection of peers. The Data Communication Module controls the streaming of data between peers. Finally, the 3D Data Retrieval Module provides access to multimedia streams and data structures necessary for visualization.

By integrating this framework with the other capabilities presented, KiReS is able to provide remotely quantitative information on the patient's movement, which includes the 3D data points of the relevant joints of the patient's skeleton. KiReS integrates a specific interface for remote interaction where local and remote video are displayed and avatars are animated with the streamed skeleton data to show real-time motion (see Figure 11). The 3D avatars on the center represent the remote (red avatar) and the local user (green avatar), respectively, and the remote and local video streams are depicted in the upper part of the interface. This interface allows the physiotherapist to interact with the patients by performing specific exercises directly in front of them. Moreover, at the same time, it makes it possible for the physiotherapist to observe the patient's movements and correct them in real time. The Teleimmersion submodule, whose technical details are explained in depth in [50], provides a reliable streaming solution for Kinect video, audio and skeleton data, being able to adapt to various network conditions by taking advantage of WebRTC multimedia streaming performance, which helps keep the latencies of audio and video within a range that guaranties an acceptable Quality of Service (QoS).

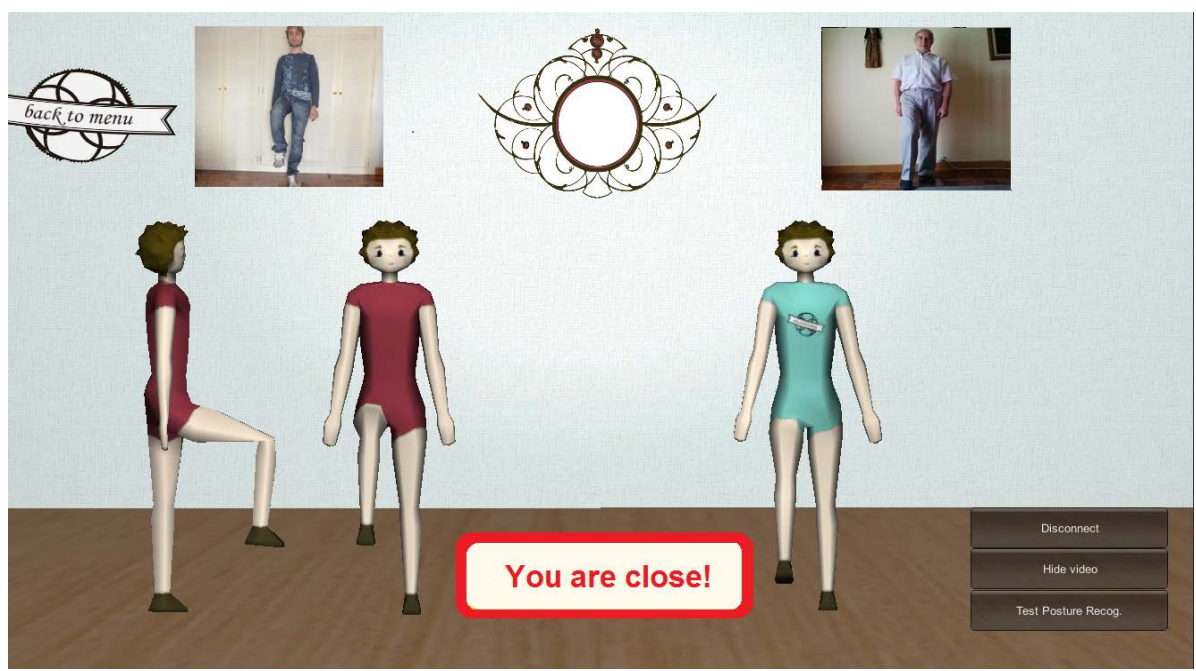

**Figure 11.** Teleimmersion in KiReS.

## 3. Results

In order to retrieve the patients' subjective perceptions, we used a Likert scale questionnaire that consisted of 13 questions about the session with five possible answers from one (strongly disagree) to five (strongly agree). The questions were divided into three categories: the system; the user experience; and the interface (see Figure 12). There was a yes/no question asking whether the patients had previously heard about telerehabilitation and also an open-ended question in which patients could write any opinion or suggestion they had about their experience with KiReS.

That questionnaire was answered by the patients that took part in two real trials after they completed each exercise session. One trial was held in a rehabilitation center in Bilbao (Spain) and the other one at Queen Elizabeth II Jubilee Hospital in Brisbane (Australia). The objective of these trials was to validate KiReS in order to evaluate the satisfaction of patients with the system. Prior to the sessions, the physiotherapists that assisted these trials designed the adequate therapy for each of the participants based on their pathologies, severity of their physical limitations and intended intensity of the recovery session. Using the KiReS tools, the physiotherapist combined exercises and established the number of series and repetitions, as well as some difficulty parameters (hold time, pain evaluation, waiting periods, etc).

Aside from the pathologies that the patients suffered, all the trials carried out during this work shared some common aspects: prior to commencing the session, the system was presented to each of the participants, and a brief explanation of the objectives and achievements of the project was given, along with a tutorial about how the system works and the elements that they were going to find in the interface during the therapy session. After that, the participants began the exercises they were assigned. Ethical clearance was provided by the relevant institutional review boards, and those participating in the study signed an informed consent form including a privacy protection statement, which was written with the endorsement of the respective institutions.

If this is your first visit, have you ever heard about telehealth or telerehabilitation?

| System | 1 | 2 | 3 | 4 | 5 |
|---|---|---|---|---|---|
| 1 This system could help with my rehabilitation. | | | | | |
| 2 This telehealth exercise session is as good as a usual exercise session. | | | | | |
| 3 I think this system would help me do my exercises at home. | | | | | |

| User experience | 1 | 2 | 3 | 4 | 5 |
|---|---|---|---|---|---|
| 4 I am satisfied with the telehealth exercise session. | | | | | |
| 5 I would like to use this system again. | | | | | |
| 6 It was easy using the system. | | | | | |
| 7 Getting used to exercising with the system was hard for me. | | | | | |
| 8 The telehealth system worked well. | | | | | |

| Interface | 1 | 2 | 3 | 4 | 5 |
|---|---|---|---|---|---|
| 9 I liked the way that the system looked. | | | | | |
| 10 The system helped me to perform the exercises. | | | | | |
| 11 It is useful to see my movements on the screen. | | | | | |
| 12 The instructions to perform the exercises helped me. | | | | | |
| 13 The system was confusing to use. | | | | | |

**Figure 12.** Questionnaire.

*3.1. Shoulder Disorder Patients (Bilbao)*

A physiotherapist from the rehabilitation center selected 11 patients that agreed to participate in a rehabilitation session. All patients suffered from shoulder disorders in only one of their arms and had been going to rehabilitation for at least one month. The ages of the patients were in a range from 32 to 58, with 45 being the average. A physiotherapist recorded a set of exercises appropriate for patients with shoulder disorders based on standard therapy protocols. This resulted in a set of 11 different exercises. These exercises were a combination of 27 postures and 16 movements (these 16 movements were also reversed, making a total of 32 movements) that the physiotherapist recorded, and using our managing tools, he combined them into the mentioned 11 exercises.

In Figure 13, we present a boxplot with the answers to each one of the 13 questions of the questionnaire (Figure 12). Each boxplot shows the mean value (red dot), the minimum value, the maximum value and the values of the interquartile ranges (IQR) Q1 (the middle value in the first half of the rank-ordered answers), Q2 (the median) and Q3 (the middle value in the second half of the rank-ordered answers). Questions of that questionnaire were classified into three categories (system, user experience and interface), and their mean values were 3.77, 3.59 and 4.05, respectively, which can be interpreted as 'quite agree' taking into account that one meant 'strongly disagree' and five 'strongly agree'. Therefore, we can say globally that the patients were moderately satisfied with the system and showed interest in using it. In the open-ended question, some of them wrote down an answer; two of the patients commented that they 'liked the system' and that it was 'a positive experience'; another one stated that 'with some adjustments it will be useful'; and one asked for 'a bigger font in the interface'. Their feedback provided new insight into how the interface and the interactions with KiReS were affecting user experience, and we used this input to further improve the system.

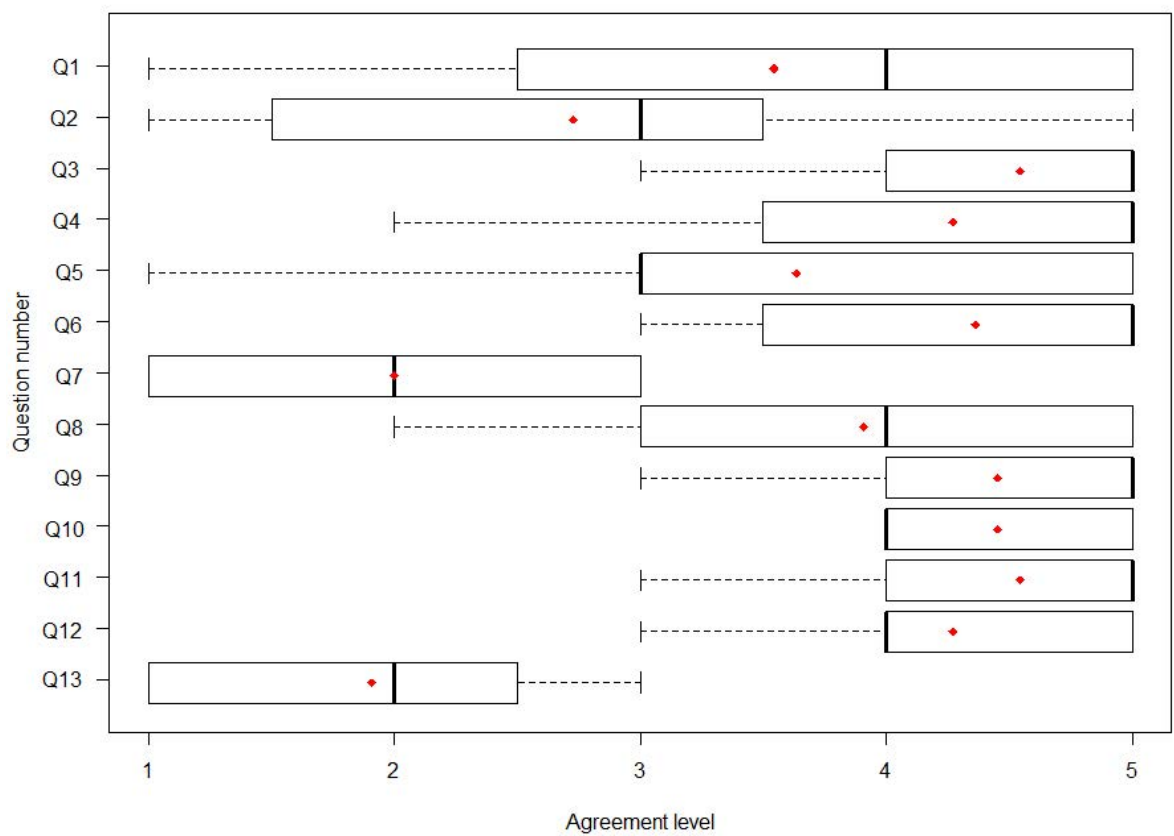

**Figure 13.** Questionnaire results from user experience at the telerehabilitation center in Bilbao: median, mean (red dot) and IQR.

Informed consent was obtained from all individual participants included in the study. The Asepeyo Medical Board also approved the study design, protocols and procedures.

*3.2. Hip Replacement Patients (Brisbane)*

In this trial, a full deployment of KiReS was made with a group of patients that had THR surgery. The inclusion criteria for the selection of the participants were: having undergone primary THR in the last four months, full weight-bearing or weight-bearing as tolerated and normal mentation. The exclusion criteria were: revision THR, restricted weight-bearing postoperatively and having co-morbidities preventing participation in a rehabilitation program. The ages of patients were in a range from 33 to 67, with 56 being the average. Most of them (five of seven) had hip replacement surgery in their left hip. Patients were invited by their treating physiotherapist to participate in the study. Nineteen questionnaires were retrieved in total from participants. None of the patients reported that they had heard about telerehabilitation or telemedicine before. Participants reported that the main negative features of the system were the size of the font and the structure of the interface, which some of them found distracting, as they considered that some of the elements were not useful.

With respect to the satisfaction results, we can mention that mean scores of 4.71 for the system and 4.4 for the user experience category were obtained (Figure 14). We also found that the evaluation of those patients who tested the system with the improved new interface (Figure 15) was higher (4.77) than with the original interface (4.43) and significantly different ($X^2$ = 6.6347, df = 2, $p$ = 0.03625).

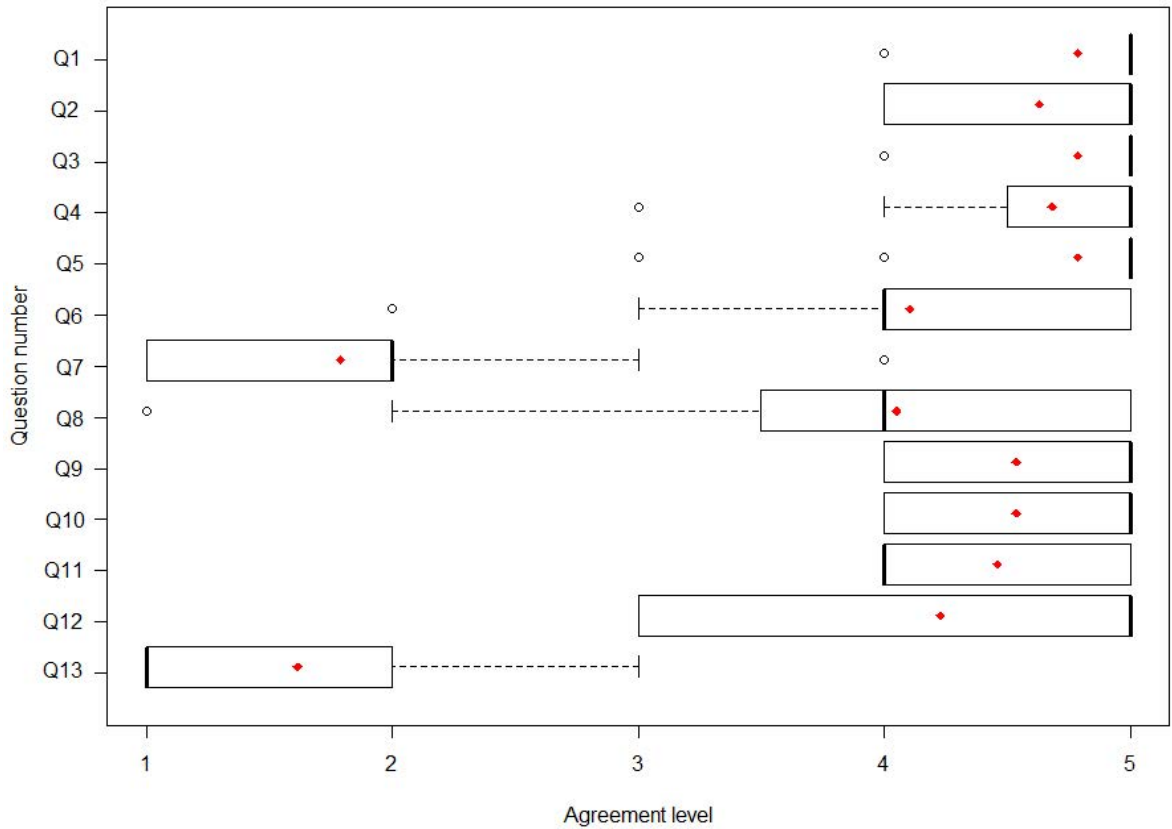

**Figure 14.** Questionnaire results from user experience at the hospital in Brisbane with the original interface: median, mean (red dot) and IQR.

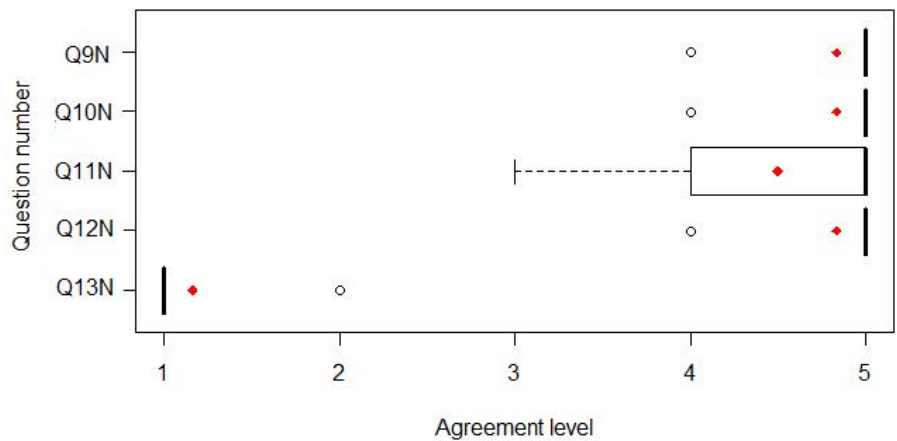

**Figure 15.** Questionnaire results from user experience at the hospital in Brisbane with the new interface: median, mean (red dot) and IQR.

Informed consent was obtained from all individual participants included in the study. This study was approved by The Office of the Metro South Human Research Ethics Committee, HREC Reference Number: HREC/13/QPAH/235, date of approval: 24 March 2014.

## 4. Discussion

There is a growing interest in developing telerehabilitation systems oriented toward the treatment of different pathologies, both physical and cognitive. Our telerehabilitation system, KiReS (Kinect TeleRehabilitation System), allows people to perform physical telerehabilitation sessions anytime in different environments. The main contribution of KiReS and what differentiates it from other systems is not only that it tackles the evaluation of the patient's evolution, but also helps physiotherapists in their daily tasks such as: storing and consulting patient records and exercises, assigning exercises, analyzing the evolution of patients and sometimes conducting two-way teleimmersion sessions. We have not found any system that provides all those functionalities altogether.

During the development of KiReS, we collaborated with physiotherapists in order to introduce adequate expert knowledge into the system, and with patients in order to validate their interest. In summary, the main goals pursued when building the KiReS system were: (1) Friendly and helpful interaction with the system: KiReS combines the use of a non-wearable motion control device with motivational interfaces based on avatars and dynamic exercise guiding, since rehabilitation depends largely on the patient's motivation and compliance to be successful. Furthermore, KiReS facilitates physiotherapists with an interface that, on the one hand, provides an easy way to define new exercises based on the therapy protocols they typically use; and on the other hand, also facilitates the task of developing tests. (2) Provision of smart data: KiReS uses different techniques to provide actionable information. It manages a novel domain ontology that provides a reference model for the representation of the physiotherapy-related information that is needed for the whole physiotherapy treatment of a patient, from when they step for the first time into the physiotherapist's office, until they are discharged. The ontology also allows the representation of patients' reports, therapy exercises, movements and evidence-based rehabilitation knowledge; and favors reasoning capabilities over therapy data for the selection of exercises and the notification of events to the therapist. This type of information is not provided by current systems, and it has been recognized as very interesting by the consulted physiotherapists. Moreover, it is able to convert low-level recorded Kinect data into qualitative measures (e.g., posture rating, exercise rating, balance, etc.) that can be useful for the physiotherapists to customize and adapt patients' therapy and for the patients to be aware of their improvements. (3) Monitoring of rehabilitation sessions: KiReS incorporates an algorithm that evaluates online performed exercises and assesses if they have been properly executed by comparing the obtained results with the recorded reference data. Automatic exercise assessment is something relevant since in home-oriented telerehabilitation systems, it is crucial that the patient is autonomously evaluated without the direct intervention of the physiotherapist during rehabilitation sessions. (4) Provision of a teleimmersion mechanism: KiReS supports a facility that allows 3D transmission of body postures and movements provided by Kinect. It facilitates stable and secure transmission of video, audio and Kinect data in real time between two peers. Thus, physiotherapists can display exercise performance remotely to the patients while also being able to observe their performance. Moreover, the patients can communicate to the physiotherapist any question or concerns about their performance. Streaming performance results showed how the combination of an open source real-time networking framework, such as WebRTC, and the Kinect camera can provide the next step in remote physical therapy with the reliable transmission of diverse medical data.

As previously mentioned, the recognition algorithm was experimentally validated by using some datasets created by five healthy volunteers. Once KiReS was operative, we tested it with real patients in two different scenarios: a rehabilitation center in Bilbao and at a hospital in Brisbane. In the first case, the system was tested with eleven patients who suffered from shoulder disorders. A physiotherapist recorded a set of exercises to be executed, and after that, the patients participated in a rehabilitation session. The accuracy of the exercise recognition algorithm was 88.14%. In the case of Brisbane, the same procedure was followed: a physiotherapist prepared the set of exercises to be performed, and seven patients that had had THR surgery participated in several rehabilitation sessions. KiReS categorized 91.88% of the exercises performed by the patients as being correct.

Finally, when we tested KiReS with patients, we also debriefed them about the system and other aspects related to telemedicine applied to their pathology. In this sense, we found that only very few of them had heard about the concept of telerehabilitation. Even though our test was oriented toward checking the functionality and usability of our telerehabilitation system and gathering the impressions of the patients, we found it relevant that the patients had neither knowledge about telerehabilitation, nor about the benefits that these systems can provide to them. Furthermore, the trials with patients showed some aspects that we consider relevant about the patients' interaction and experience with the system. First, we found that the interaction with the Kinect was easy to learn for the patients and that they thought the system comfortable to interact with. Second, they perceived the system as a useful complement to their regular therapy sessions, which can enhance healthcare assistance. However, they considered it less effective than ordinary sessions. This is nevertheless the objective of telerehabilitation, to be complementary to traditional therapy, making it more accessible, but without replacing the traditional rehabilitation. Third, the patients showed interest in using the system again and manifested being satisfied with the experience. Finally, patients found the 3D avatars a helpful source of information, and they rated the interface and interaction with the system in a positive way. In summary, the trials showed that the system can provide benefits for the patients and the interest they have in this kind of technology, but new studies in which larger populations would participate are needed to find the best balance between traditional rehabilitation and telerehabilitation so that the results and user experience with the system can keep improving.

*Limitations*

Though it was announced that the Kinect device itself was discontinued, Microsoft still provides support for Kinect SDK developers [51]. Furthermore, according to Microsoft, "Microsoft is working with Intel to provide an option for developers looking to transition from the Kinect for Windows platform". Intel RealSense cameras [52] or Orbbec cameras [53] are an alternative to the Kinect, as they provide similar features.

From the point of view of our system, even though our current implementation relies on the Kinect for motion tracking, other depth camera devices such as those mentioned above could be integrated with it. Skeleton tracking is the main requirement for a 3D camera to be compatible with the presented system. As the system is designed to be modular, updating the Joints Data Processing module (shown in Figure 2) would be the main change necessary for such an integration. Other modules would suffer only minimal changes. For instance, the Recognition and Evaluation and the Knowledge Extraction Modules would a priori only require a new mapping of variables to fit the skeleton provided by a different tracking device. The interface would probably require an update to accommodate the new device's SDK. In our experience, however, a depth camera's SDK usually provides a similar framework to give access to color images, depth images, etc. The Teleimmersion Module would not have to change if joint data and depth images were stored in the local database (see Figure 2) by using the same format as before. Therefore, integration in this respect would be straightforward.

## 5. Conclusions

The goal of the system presented in this paper was to go a step further in the development of telerehabilitation systems and to show how new relevant functionalities can be incorporated with them, which can serve as a great help to patients and physiotherapists. Thus, the system incorporates methods that allow creating new exercises and tests in a friendly way; helping physiotherapists in the task of assigning the adequate exercises and in the task of evaluating the evolution of patients; empowering patients in their rehabilitation process, allowing them to perform exercises in an autonomous way and providing immediate feedback about how they are performing them; and finally, to perform the exercises in real time, remotely, in a virtual environment.

The results obtained so far using KiReS show its suitability for telerehabilitation and a good quality user experience. The patients who used it found that the interaction with it was friendly; they considered it as a complement to their therapy that can improve medical attention; and they showed a predisposition to using the system again. KiReS can be extended and as future work we plan to enhance the information KiReS retrieves by adding bio-signal tracking devices. Thus, it would be possible to extend the reasoning and data analysis capabilities of the system with these new inputs.

**Author Contributions:** All authors participated in the definition and design of the system and in the discussion of its relevant aspects. D.A., A.G. and A.I. were involved in the development and the deployment of the telerehabilitation system. I.B. and J.B. designed and implemented the ontology. D.A. designed the experiments and was involved in data collection and interpretation during the trials. All authors wrote the manuscript and read and approved the final manuscript.

**Funding:** This research was funded by the Spanish Ministry of Economy and Competitiveness grant number FEDER/TIN2016-78011-C4-2R.

**Acknowledgments:** Authors thank Jon Torres-Unda and Jesús Seco for their valuable collaboration with physiotherapy-related aspects and for their feedback about KiReS. Some parts of this paper are available in an earlier PhD thesis publication, accessible in the following repository of the University of the Basque Country UPV/EHU: https://addi.ehu.es/handle/10810/16068.

**Conflicts of Interest:** The authors declare no conflict of interest. The founding sponsors had no role in the design of the study; in the collection, analyses or interpretation of data; in the writing of the manuscript; nor in the decision to publish the results.